# Deformation Potential Carrier-Phonon Scattering in Semiconducting Carbon Nanotube Transistors

G. Pennington[1], A. Akturk[1], N. Goldsman[1], and A. Wickenden[2]
[1]Department of Electrical Engineering, University of Maryland, College Park, MD 20742, USA
[2]U.S. Army Research Laboratory, 2800 Powder Mill Road, Adelphi, MD 20783, USA
email: (garyp, aturka, neil)@glue.umd.edu,  awickenden@arl.army.mil

**Abstract:** Theoretical calculations of carrier transport in single-walled carbon nanotubes are compared with recent experiments. Carrier-phonon scattering is accounted for using the deformation potential approximation. Comparing with experiments, a deformation potential coupling constant of 14eV is determined for semiconducting carbon nanotubes. Theory is shown to closely predict the low-field mobility, on conductance, and on resistance of field-effect transistors as a function of induced nanotube charge density, diameter, and temperature. Results indicate that the device conductance is reduced as multiple subband channels conduct due to strong intersubband scattering. Comparison with experiment allows identification of the mean free path ($L_m$) in semiconducting carbon nanotubes. As the device turns on, $L_m$ is found to increase significantly. When the device is in the on state, the mean free path ($L_{m\text{-}ON}$) varies linearly with tube diameter and inversely with temperature. Intersubband scattering is found to strongly decrease $L_{m\text{-}ON}$ when a few subbands are occupied. When 3 subband channels are considered at room temperature, $L_{m\text{-}ON}$ decreases from 570nm to 200nm for a 4nm diameter tube when intersubband scattering is included. Since the subband spacing increases with decreasing tube diameter, the effects of intersubband scattering are reduced for smaller diameters.

Carbon nanotube field-effect transistors (CNT-FET) have many potential applications in electronics.[1-3] Many experimental[4-8] and theoretical[9-19] studies indicate that phonon scattering significantly affects the transport properties of carriers in CNTs. While the effects of scatters such as lattice defects can be potentially reduced in synthesis, phonon scattering is intrinsic to the nanotube and determines the performance limits of the device. A full understanding of phonon-limited transport is therefore very important for the successful design of CNT-FETs. In this letter, the carrier long-wavelength phonon interaction is represented using the deformation potential approximation. This approximation has been very successful in modeling phonon scattering in conventional FETs[20], and is expected to be applicable to nanotubes.[9-15] A recent experiment[5], reporting semiconducting CNT-FET device characteristics with varying carrier density (N), temperature (T), and tube diameter (d), now allows a comparison with theory. Since the devices studied where long (L = 4-15μm) and relatively free of defects, they are ideally suited for comparison with semiclassical transport theory. We show here that solution of the Boltzmann transport equation, involving a relatively simple carrier-phonon scattering theory in the collision integral, agrees very well with these experiments. This allows identification of the deformation potential in semiconducting carbon nanotubes.

Carrier subbands and phonon subbranches in a nanotube can be specified by a circumferential wavevector β, with β=2/3d, 4/3d, and 8/3d for the first 3 subbands.[21-22] Wavevectors k and q along the tube axis further define the carrier and phonon eigenstates respectively. Using Fermi's golden rule, the phonon mediated scattering rate from initial carrier state (k,β) to final carrier state (k+q,β+Δβ) is given by [10-11]

$$W_{k,\beta \to k+q,\beta+\Delta\beta} = \frac{\hbar D^2(q,\Delta\beta) \cdot Dos(k+q,\beta+\Delta\beta)}{\rho_{2D} E_P d}\left[N_P(E_P) + \frac{1}{2}(\pm)\frac{1}{2}\right]. \tag{1}$$

Here D, $E_P$, and $N_P$ are the deformation potential per unit lattice displacement, phonon energy, and phonon number respectively. $N_P$ is determined by the Bose-Einstein equilibrium distribution. The mass density of the nanotube is $\rho_{2D} \cdot \pi d$, where $\rho_{2D}=7.6 \times 10^{-7}$ kg/m$^2$. The density of carrier states is DOS(k,β) = $\xi/\hbar V_G$, where $\xi=(1+\beta^2/k^2)^{1/2}$ [21-22] and $V_G=9 \times 10^5$ m/s is the Fermi velocity of graphene. Here DOS(k+q,β+Δ β) is equal to 1/4 of the total density of final carrier states at k+q due to conservation of both spin and circumferential momentum in the interaction.

Focusing on low-field degenerate conditions in this work, scattering with low energy phonons is important. For scattering amongst 3 subbands, the 3 acoustic phonon branches in graphene can be represented in the nanotube as the 9 subbranches in Fig. 1. Here the spectrum, $E_P$ in Eq. (1), is determined using continuum modeling.[9] As shown, $E_P \cdot d$ vs. $q \cdot d$ is independent of tube diameter, and the spectrum in Fig. 1 can be used for any single-walled CNT. We will not consider scattering between degenerate subband valleys [10] here since the phonon energies involved are too large to be significant under degenerate conditions.

The deformation potential can be found from the contraction/dilation tensor in the long wavelength limit.[9][14][23-24] The diagonal ($D_1$) and off-diagonal ($D_2$) elements are:

$$D_1 = g_1\left[\frac{2}{d} \cdot \frac{E_P^2(q=0)}{E_{PB}^2} + 2q\frac{V_T^2}{V_L^2}\right], \quad D_{2S} = 2qg_2\frac{V_T^2}{V_L^2}\cos(3\varphi_c), \quad D_{2T} = qg_2\sin(3\varphi_c) \tag{2}$$

Here the longitudinal and transverse sound velocities in graphene are $V_L$=21.1 km/s and $V_T$=15 km/s. Coupling constants are $g_1$ and $g_2$, $\varphi_c$ is the chiral angle, and $E_{PB}$ is the energy of the breathing mode in Fig. 1. It's expected that g1 can be approximated by the in-plane acoustic deformation potential of graphene ($g_1 \sim 16$eV).[9][25] The diagonal term $D_1$ contains terms of zero and 1$^{st}$ order in q. For the S, $W_1$, and $W_2$ modes only the 1$^{st}$ order term is used. An exception to Eq. (2) occurs in the case the twisting mode (T) where $D_1$=0 since this mode does not couple to axial or radial vibrations. Predictions for $g_2$ vary from 3γ[23-24] to γ/2[9], where γ~3eV is the nearest neighbor tight-binding interaction energy. Since it is expected that $D_1 \gg D_2$, the diagonal component of D enters the scattering rate in Eq. (1) unless the Fermi energy is very close to the Fermi point of graphene. In this case, $D_1$ vanishes.[9] This special case is expected for low-energy carriers in metallic nanotubes when only the 1$^{st}$ subband is occupied. Carriers will therefore scatter only with stretching (S) and twisting (T) modes under these conditions. The off-diagonal components for these modes, $D_{2S}$ and $D_{2T}$ respectively in Eq. (2), depend on the chiral angle of the nanotube. It is still unclear whether the diagonal or off-diagonal deformation potential is applicable for the S and T modes in semiconducting nanotubes[9][14][23-24], both cases are studied here.

We investigate carrier transport in single-walled semiconducting carbon nanotubes by solving the Boltzmann equation. Such a treatment is applicable when the length of the nanotube is much larger than the mean free path for scattering. Previous work in nanotubes has focused on solutions of the Boltzmann transport equation by Monte Carlo[10-12] or by direct iteration[16]. We employ the later method here since solution by iterative methods is well suited for near-equilibrium and degenerate conditions applicable in low-field CNT-FET applications.[4-5] We find the carrier distribution function $f_{k,\beta}$ by solving the spatially homogenous multisubband Boltzmann transport equation[16][26-27]

$$\frac{eF}{\hbar}\frac{\partial f_{k,\beta}}{\partial k} = \sum_{q,\Delta\beta}\left[f_{k+q,\beta+\Delta\beta}(1-f_{k,\beta})\cdot W_{k+q,\beta+\Delta\beta \to k,\beta} - f_{k,\beta}(1-f_{k+q,\beta+\Delta\beta})\cdot W_{k,\beta \to k+q,\beta+\Delta\beta}\right], \quad (3)$$

for a small field F (FL<<$k_B$T) directed along the tube axis. The distribution function ($f_{k\beta}$) is solved for a given Fermi energy through specification of the 1-D charge density N. The conductance is then $G(N)=\Sigma_\beta G_\beta=eN_\beta\mu_\beta(N)/L$, where the sum rule is used to sum over each subband (defined by unique β). The mobility $\mu_\beta$ is found from the drift velocity ($V_{d\beta}$) and field according to

$$\mu_\beta = \frac{V_{d\beta}}{F} = \frac{V_G}{F}\cdot\frac{\sum_k f_{k,\beta}/\xi_\beta \cdot \frac{k}{|k|}}{\sum_k f_{k,\beta}}, \quad \mu_{FE} = \frac{\partial(N_\beta\mu_\beta)}{\partial N}, \quad (4)$$

The field-effect mobility for the multisubband system is $\mu_{EF}$. Fig. 2 shows theoretical results for the room temperature conductance of two carbon nanotubes with varying diameter. Here we use the diagonal component of the deformation potential ($D_1$ only) with $g_1$=14eV. As N increases, the conductance initially rises as the 1$^{st}$ subband is populated. If no other subbands are considered, $G_{2/3d}(N)$ (β=2/3d for subband 1) saturates once the density of carrier states in the 1$^{st}$ subband becomes nearly constant.[5] Both the initial slope and the saturation level of the conductance increase with tube diameter. When multiple subbands are included, intersubband scattering is found to reduce G even though more conducting channels are opened up with increasing N. Here the conductance of a given subband channel is dramatically reduced when the Fermi level is near the minimum of a higher subband. At this point DOS(EF) is much larger in the higher subband. Scattering into the higher subband greatly exceeds out-scattering whereas for the lower subband out-scattering greatly exceeds in-scattering. This produces valleys in $G_\beta$ and subsequently in the multisubband conductance in Fig. 2. Since the conductance can also be represented as $G_\beta=4e^2L_\beta/hL$, results in Fig. 2 also show the mean free path $L_\beta$ in microns. The mean free path of the multisubband system would be $L_m=\Sigma_\beta L_\beta/N_{SB}$, where $N_{SB}$ is the number of subbands included. Therefore, the preceding discussion for the density dependent conductance also applies for $L_m(N)$. The semiclassical Boltzmann transport treatment would be applicable when L>10·$L_m$. Since $L_m$ rises with N, transport in a nanotube FET may be semiclassical at low N where $L_m$ is small, but not when the on conductance is reached.

It is important to point out that the mobility in Eq. (4) is essentially normalized by the total charge density N~$\Sigma f_{k\beta}$, including both mobile and non-mobile carriers. If the fraction of non-mobile carriers increases, μ will tend to be reduced. The correct conductance due to only mobile charge is still obtained since G is proportional to Nμ. Mobilities, corresponding to results for G in Fig. 2, are shown in Fig. 3. Both μ and $\mu_{FE}$ peak at low carrier density where a larger percentage of carriers are in mobile states. Here the Fermi level is below the 1$^{st}$ subband and mobile carriers represent the tail of the distribution function. As the Fermi level enters the subband structure with increasing gate-induced charge density N, the mobility decreases since many carrier states below the Fermi level are non-conducting. The peak values of the mobility are found to increase with increasing tube diameter (~$d^2$) as in previous work.[5][10][16] Experimental measurements of the field-effect mobility in single-walled carbon nanotubes[4-5] allows a comparison with theory. Since $L_m$ is small when the mobility peaks, semiclassical transport theory has increased applicability. Results are shown in Fig. 4, where $\mu_{FE}$ peak values as a function of tube diameter are well reproduced by theory when the diagonal component of the deformation potential is used with $g_1$=14eV. However, experiments do not compare well when the off-diagonal component is used for the S and T modes, even when $g_2$=9eV.

In Fig. 5 we compare the maximum conductance with results from experiments.[5] Results also indicate the mean free path ($L_{m-ON}$) and $L/R_{NT-ON}$, where $R_{NT-ON}$ is the on resistance of the nanotube when the FET is turned on. Result show that intersubband scattering reduces $L_{m-ON}$ significantly. Also shown is the fit to conductance peak measurements in ref. [5] using a fitting parameter α.=18.4 m/Ks. Theoretical predictions agree with experiments showing that the on resistance of the nanotube varies as ~1/d, whereas the mean free path of the nanotube varies as ~d. Theoretical predictions for the temperature-dependent mobility peak and the on resistance are compared with experiments [5] in Fig. 6. Here we have included a temperature-independent contact resistance $R_C$=28kΩ as measured in ref. [5]. Theory predicts that the $μ_{FE}$ peak varies as ~1/T, as in previous studies [14][16], whereas $R_{ON}$ increases linearly with increasing temperature. As seen in Fig. 6, theory and experiment are in close agreement.

In conclusion, we have compared theoretical results based on a solution of the Boltzmann transport equation with experiments on single-walled semiconducting CNT-FETs.[5] Carriers occupy up to 3 subbands and are subject to deformation potential scattering with low energy phonons. We find close agreement between theory and experiment when the diagonal component of the deformation potential [9] is considered with a coupling constant of 14eV. Predictions for the electronic properties of semiconducting CNT-FET devices in degenerate systems closely follow experimental findings [5]: $μ_{FE}$ ~$d^2$/T, and when the transistor is on $G_{ON}$~d/T, $L_{m-ON}$~d/T, and $R_{ON}$~T/d. Intersubband scattering is found to significantly reduce $L_{m-ON}$ in large diameter tubes.

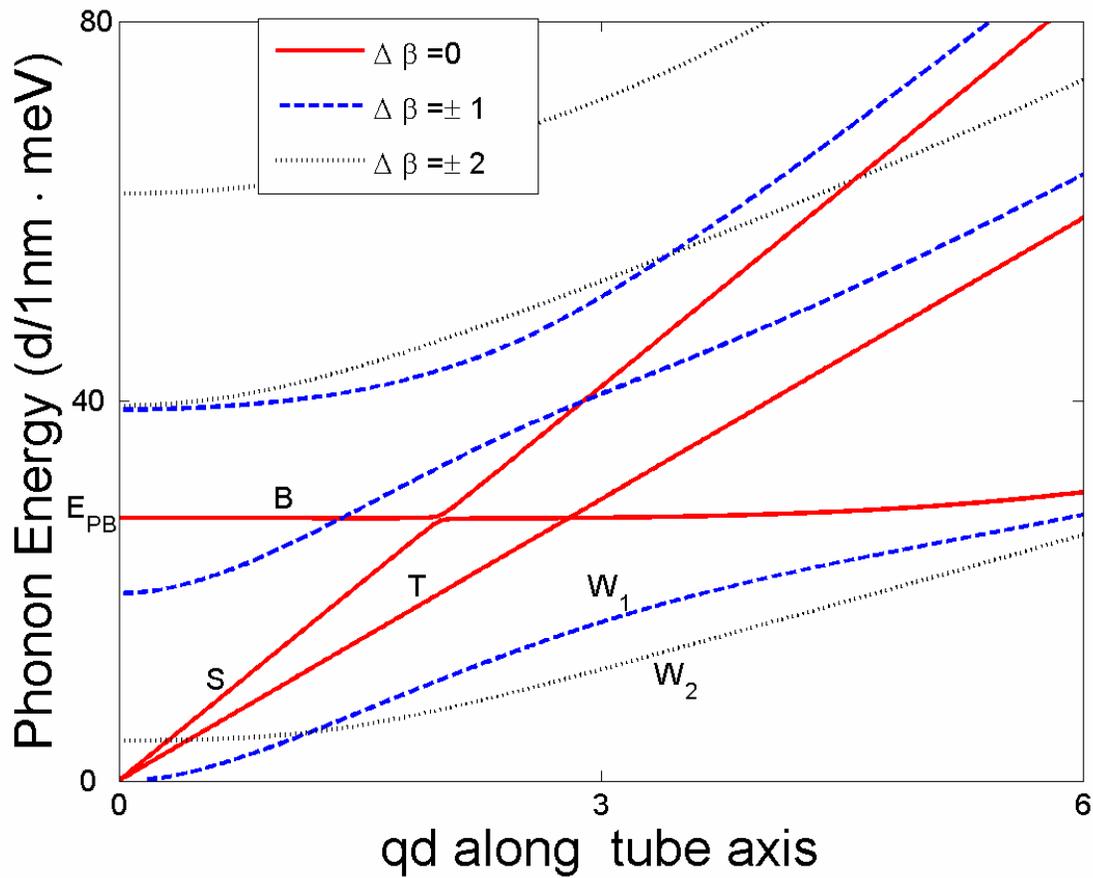

Figure 1: Continuum model results for the low-energy phonon energy ($E_P$) vs. wavevector (q) for a single-walled carbon nanotube. Subbranches for intrasubband scattering correspond to stretching (S), twisting (T) and breathing (B) modes. Subbranches for scattering between carrier subbands 1-3 are also shown. As both $E_P$ and q are normalized by the diameter (d), results are valid for any d. A factor of (d/1nm -1) must be added to the continuum results for the $W_2$ subbranch for universality.

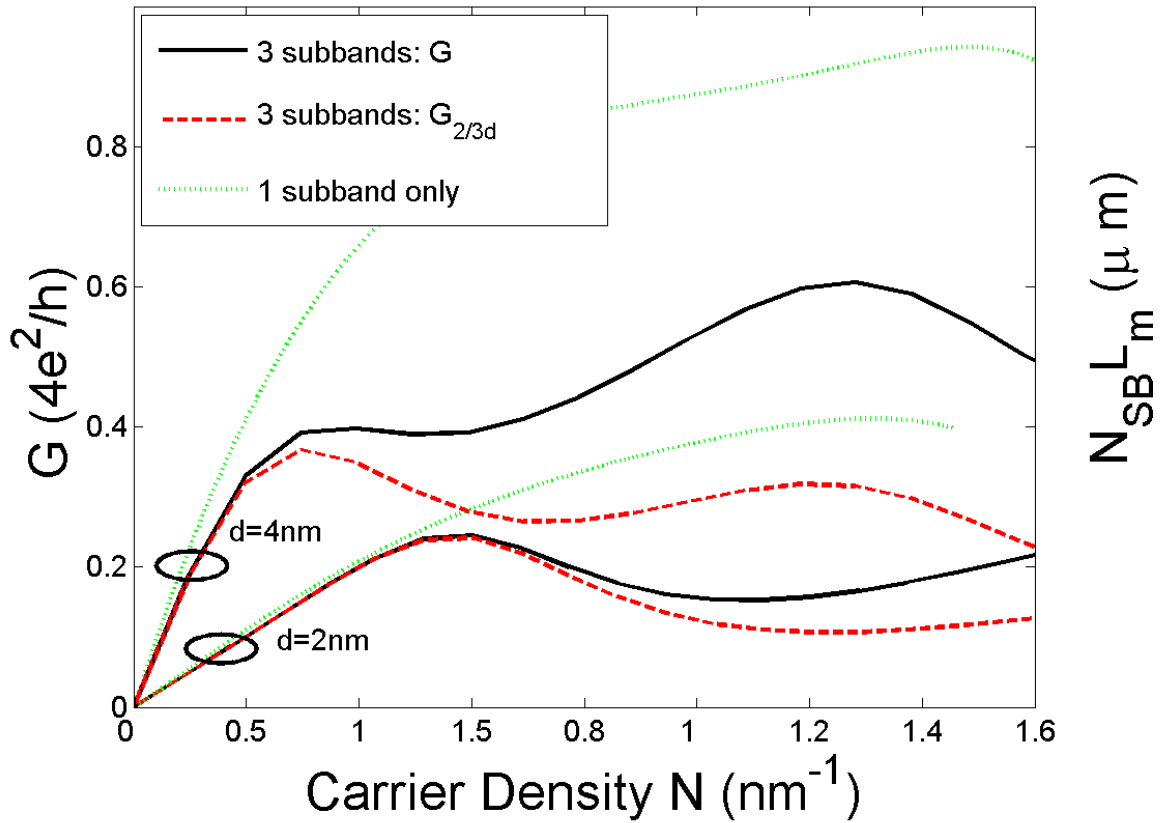

Figure 2: Theoretical room temperature conductance of a 1 micron single-walled carbon nanotube as a function of charge density. Diameters of 4nm and 2nm are shown. For each tube the total conductance $G=\sum_\beta G_\beta$ and $G_{2/3d}$ for only subband one are shown when all 3 subbands are considered. Also shown is the case when only one subband is considered ($G=G_{2/3d}$ here). Results also give the mean free path $L_m$ in microns multiplied by the number of subbands included $N_{SB}$ since the tube length is 1 μm.

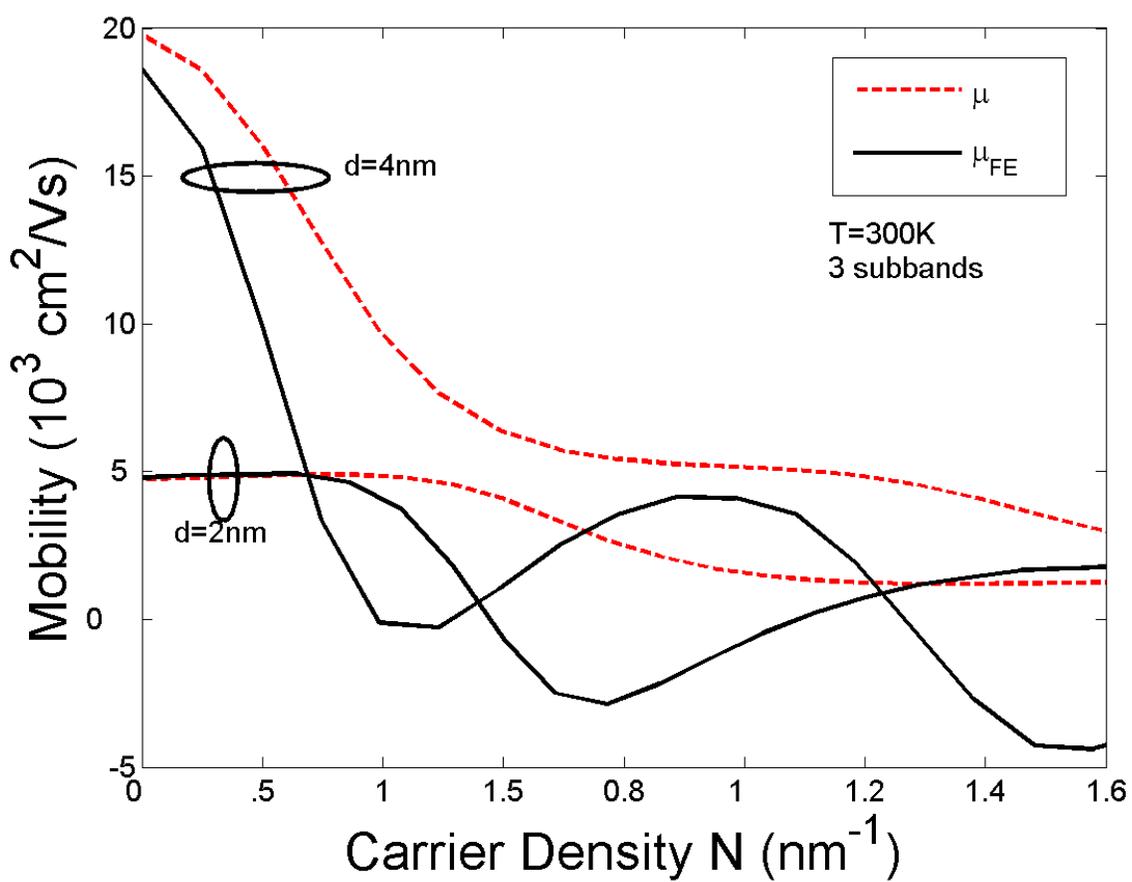

Figure 3: Theoretical mobility vs. charge density for single-walled carbon nanotubes with diameters 4nm and 2 nm. Results are for 300K and 3 subbands are used in the theory.

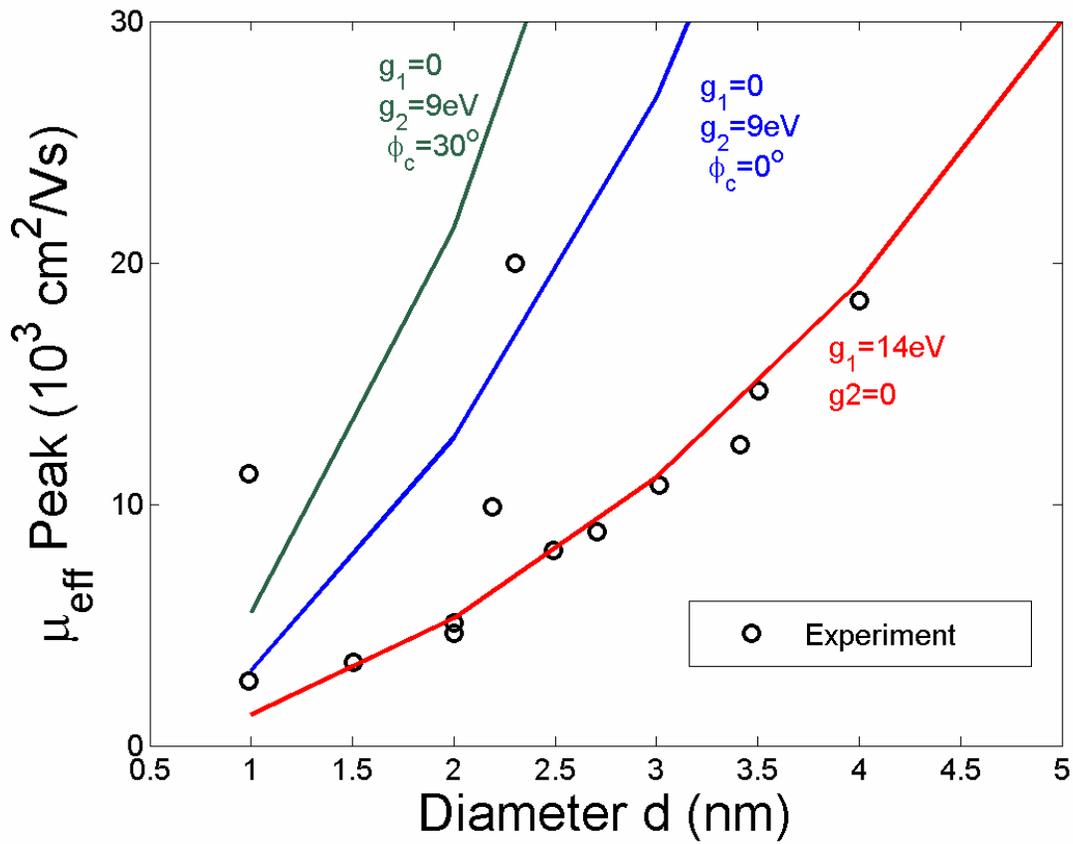

Figure 4: Comparison between experimental [5] and theoretical field-effect mobility peaks as a function of single-walled nanotube diameter. Deformation potential $D_1$ in Eq. (2) agrees well with measurements whereas $D_2$ does not. Results are at 300K.

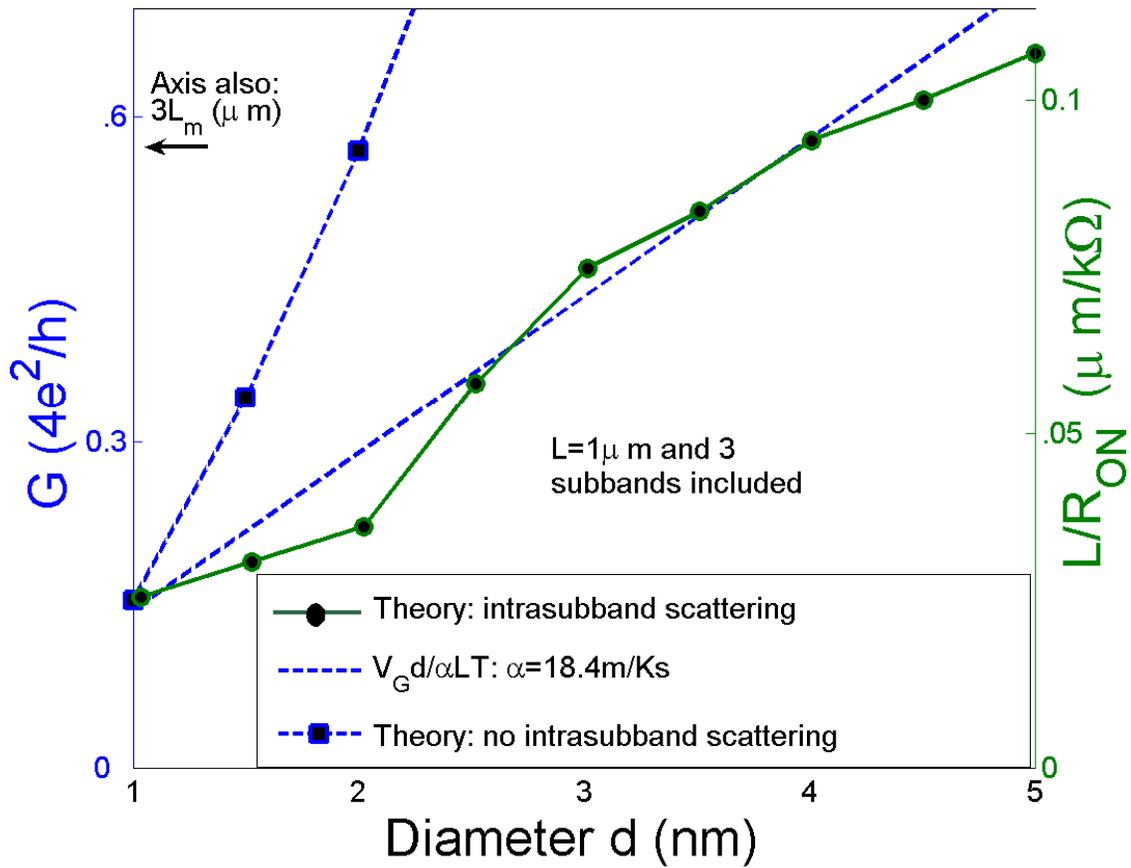

Figure 5: Simulations of the theoretical peak conductance for a one micron length single-walled carbon nanotube as a function of diameter at T=300K. Results also indicate the on state mean free path $L_{m-ON}(d)$ and the on resistance of the nanotube $R_{CNT-ON}$ (excluding contacts). Intersubband scattering strongly reduces the on conductance and mean free path for larger diameter tubes.

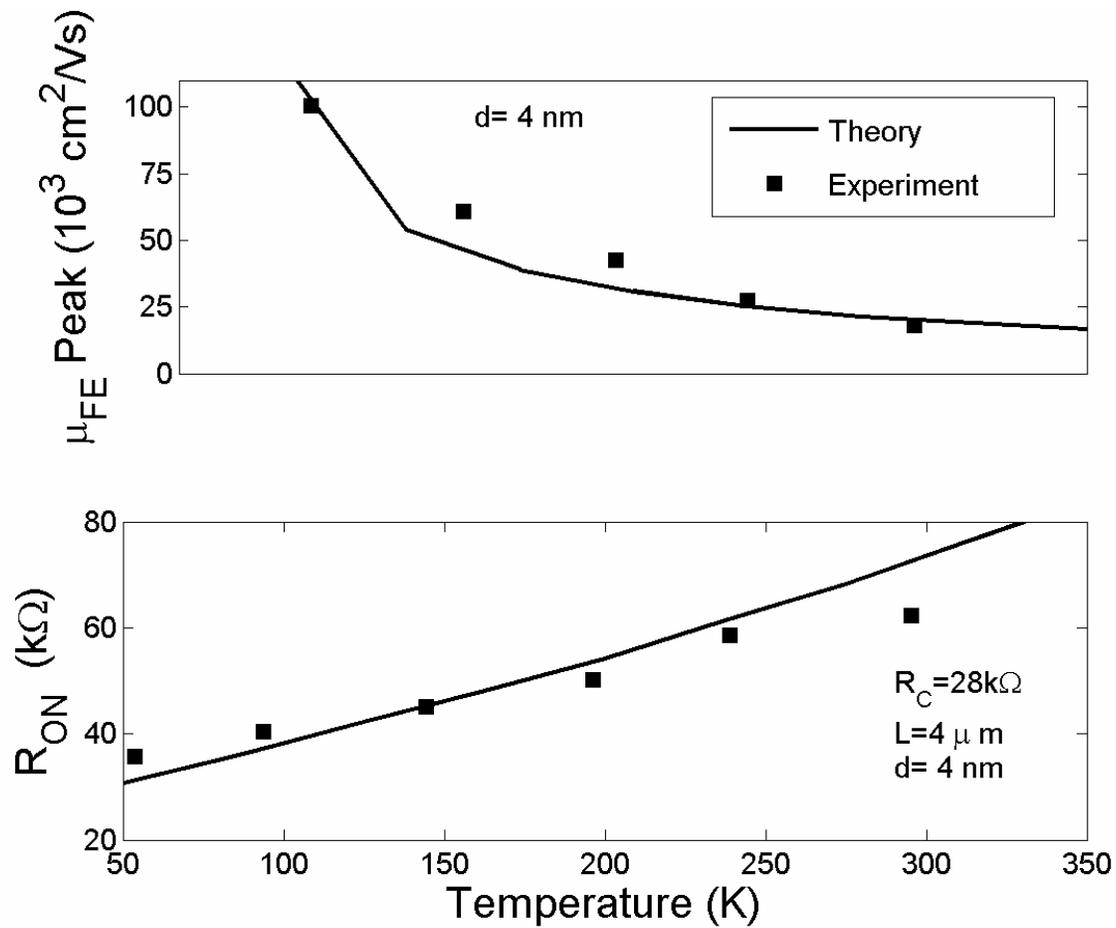

Figure 6: Comparison of theoretical and experimental [5] peak field-effect mobility and on resistance vs. temperature. A single-walled nanotube of length L=4 microns and a diameter of d=4nm is considered. The on resistance is $R_C + R_{CNT\text{-}ON}$, where $R_C$ is 28kΩ.[5]